\documentclass{elsart1p}
\usepackage{graphicx,amssymb}
\usepackage[%
  breaklinks=true,%
  colorlinks=true,%
  linkcolor=blue,anchorcolor=blue,%
  citecolor=blue,filecolor=blue,%
  menucolor=blue,pagecolor=blue,%
  urlcolor=blue]{hyperref}

\makeatletter
\def\elsartstyle{%
    \def\normalsize{\@setfontsize\normalsize\@xiipt{14.5}}
    \def\small{\@setfontsize\small\@xipt{13.6}}
    \let\footnotesize=\small
    \def\large{\@setfontsize\large\@xivpt{18}}
    \def\Large{\@setfontsize\Large\@xviipt{22}}
    \skip\@mpfootins = 18\p@ \@plus 2\p@
    \normalsize
}
\@ifundefined{square}{}{}
\makeatother

\begin{document}
 
\begin{frontmatter}
\title{Object detection in multi-epoch data}
 
\author{G. Jogesh Babu\thanksref{nsf}}
\address{Department of Statistics,
326 Joab L. Thomas Building,
The Pennsylvania State University,
University Park, PA 16802-2111, USA}
\ead{babu@stat.psu.edu}
\ead[url]{www.stat.psu.edu/$\sim$babu}
\thanks[nsf]{Supported in part by the National Science
Foundation grants AST-0434234 and AST-0326524}
\author{A. Mahabal\thanksref{grist}},
\ead{aam@astro.caltech.edu}
\author{S. G. Djorgovski\thanksref{grist}}
\address{Division of Physics, Mathematics, and Astronomy,
California Institute of Technology,
Pasadena, CA 91125, USA}
\ead{george@astro.caltech.edu}
\author{R. Williams\thanksref{grist}}
\ead{roy@caltech.edu}
\address{Center for Advanced Computing Research,
California Institute of Technology,
Pasadena, CA 91125, USA}
\thanks[grist]{Supported in part by the National Science
Foundation grants AST-0326524 and AST-0407448}
\begin{abstract}
In astronomy multiple images are frequently obtained at the same position of the sky for follow-up co-addition as it helps one go deeper and look for fainter objects. With
large scale panchromatic synoptic surveys becoming more common, image co-addition has become even more necessary as new observations start to get compared with co-added fiducial sky in real time. The standard co-addition techniques have included straight averages, variance weighted averages, medians etc. A more sophisticated nonlinear response chi-square method is also used when it is known that the data are background noise limited and the point spread function is homogenized in all channels.
A more robust object detection technique capable of detecting faint sources, even those not seen at all epochs which will normally be smoothed out in traditional methods, is described. The analysis at each pixel level is based on a formula similar to {\bf Mahalanobis distance}. The method does not depend on the point spread function.
\end{abstract}
 
\end{frontmatter}

\section{Introduction}\label{sec.object}
 
Many major projects, ongoing and future synoptic surveys,
MACHO, OGLE, Palomar-QUEST, Pan-STARRS and LSST, involve
repeated scans of large areas of the sky in several spectral
bands. Thus an important area of recent astronomical research has been the
investigation of source detection in multi-epoch data. A question
frequently asked is: What is the best way to
combine image regions with low signal to detect faint objects?
Historically two basic methods have been used to search for faint
astronomical objects:
\begin{enumerate}
\item Use a larger telescope to collect a larger number of photons
\item Stack a large number of registered images in order to improve
the signal-to-noise ratio.
\end{enumerate}
The former has
engineering and monetary limitations while the later may not
work for transients that are seen only once and for very faint objects
where the signal remains below detection threshold even after the
pixel-to-pixel coadding of several images.
Szalay et al. (1999) had proposed a method that used chi-square
coaddition by treating adjacent pixels to be uncorrelated.
It is an improvement
over standard coaddition, but since the pixels
are in fact correlated, it has its own limitations.
The procedure described here is a more robust technique to detect
faint sources from multi-epoch data. It is designed not only to have high
sensitivity, but to detect changes between images.

The analysis at each pixel level
is based on a statistic similar to the measure of distance 
introduced a by P. C. Mahalanobis in 1936 (Johnson \& Wichern 1992, and Atkinson et al.\ 2004). The {\bf Mahalanobis distance} $D_M$ from a group
of values with mean vector $\nu$ and covariance matrix $\Lambda$ from a
multivariate vector {\bf x} is defined as
$$ D_M({\bf x}) = \sqrt{({\bf x} - \nu)\Lambda^{-1}({\bf x} - \nu)^T}.$$
It is used in classical multivariate analysis and differs from Euclidean distance. It is scale-invariant, and is based on correlations between variables by which different patterns can be identified and analyzed. Mahalanobis distance can also be defined as dissimilarity measure 
$$d_m({\bf X}, {\bf Y}) = \sqrt{({\bf X}-{\bf Y})\Lambda^{-1}({\bf X}-{\bf Y})^T}$$
between two random vectors {\bf X} and {\bf Y} of the same distribution with 
covariance matrix $\Lambda$. If the covariance matrix is identity matrix,
then the Mahalanobis distance reduces to the Euclidean distance.
It is a useful way of determining similarity of an unknown sample set to a known one (Banks et al.\ 2004, and Atkinson et al.\ 2004). Mahalanobis distance is available in R Stats package.

The method does not depend on the
point spread function, requiring only that the pixel size is invariant.
In the following sections we provide the details of the method as well as the tests we have
run so far and future plans for large scale implementation.

 
\section{Methodology}\label{method}
 
We start with $N$ images of a given region of the sky. We first use standard
techniques to ensure that all images are the same size in terms of area
covered as well as the pixel dimensions.
The images are also background subtracted so the mean value in empty parts
of each image is close to zero. Suppose the pixel dimensions of
each of the images is $(m,n)$. So there are $R$ pixels in each of
the image ($=m\times n$).  We
concentrate on a single pixel $r$. Let {\bf f}$^r=(f^r_1, \dots,
f^r_N)$ denote the row vector of photon counts in $N$ images at
pixel $r$. We assume that {\bf f}$^r$ is approximately
Gaussian, {\it i.e.}
    $$ {\bf f}^r \sim MVN({\bf M}, \Sigma),$$
{\bf M}$=(\mu_1, \dots, \mu_N)$ is the vector of means
and  $\Sigma$ is the variance-covariance matrix. Then ${\bf X}=({\bf
f}^r - {\bf M})\Sigma^{-1}({\bf f}^r - {\bf M})^T$ has approximately
$\chi^2$ distribution. Since approximations are involved, use of bootstrap
method, instead of using $\chi^2$ tables, to find the critical values (thresholds)
at various significance levels is better. Bootstrapping such quadratic forms
require special care, see Babu (1984) for details.
 
This method takes into account potential
information in different wave bands. The probability of detecting a
source when $ {\bf X} > y $ can be obtained from
$\chi^2$ tables. This incorporates potential correlations of
background in different images. If the $f^r_i$ are all uncorrelated,
then the off-diagonal entries of $\Sigma$ are all zero.
 
Means and covariances can be estimated in two different ways. If the
background $M_i$ is spatially constant in the i-th image, the
covariance matrix $\Sigma$ has entries  $s_{i,k}= (1/R)\sum (f^r_i
-M_i)(f^r_k - M_k)$ where $R$ is the number of pixels in each image
or cutout. Now the $i,k$-th entry of covariance
matrix $\Sigma$ is $s_{i,k}$. This is applicable if the background
is similar throughout the image. For spatially variable background,
we currently use 9
pixels (the pixel under question and its 8 neighbors)
to estimate the covariance matrix $\Sigma(r)$ and the mean
vector $M(r)$. If there is no source at pixel r, then $(f^r-
M(r))\Sigma(r)^{-1} (f^r- M(r))^T$ has approximately chi-square
distribution.
 
Standard programs like sextractor (SourceExtractor,
available from http://terapix.iap.fr) look for a
connected set of pixels above a threshold resulting in a list of
astronomical objects (called catalog in astronomy jargon) with various
flags indicating its validity as an astronomical object based
on prior knowledge like that of
shapes and profiles of objects built in to the program.
 
While this technique may not do any better than usual methods
for bright objects,
under lower signal conditions it is expected to strongly
outperform the traditional methods.
We will extensively test our technique on the already available
PQ (Palomar-QUEST) data in readiness for the even larger stream of
LSST data.
 
As we are testing multiple hypotheses, we need to avoid too many
false positives. When there are large number of pixels above the threshold, there may as well be too many falsely discovered source pixels. The proportion of falsely discovered source pixels should be kept to a minimum with any source detection algorithm. The method introduced by Benjamini \& Hochberg (1995) (see also Benjamini \& Hochberg 1997, and Benjamini \& Yekutieli 2005), called {\it False Discovery Rate} (FDR) does precisely this, controlling the proportion of incorrectly rejected null hypotheses. T
he method allows {\it a priori} specification of proportion of false discoveries to total discoveries, and the procedure is independent of the source distribution. An implementation of this procedure is discussed in Hopkins et al.\ (2002).
However, in the examples discussed in the current article, as the detected sources are few and far apart, we shall consider FDR implementation for our method elsewhere.
Babu (2004) developed results on bootstrapped
empirical process that are relevant in the study of FDR problems.
 
So far we modeled based on Gaussian assumptions, where least squares methods and maximum likelihood methods coincide. Further, the dependency of pixel data across different filters can be modeled and analyzed with minimum difficulty through covariances and joint multivariate distributions.  For Poisson data, modeling the multi-dimensional data with statistical dependence structure is far from clear. Instead it is more natural to consider modeling functional dependency of Poisson rates across filters. This w
ill be discussed in a later article.

\section{Tests}\label{tests}
 
We carried out a series of tests using artificial data as well as
{\it B R I} images from the
Palomar-QUEST survey (http://palquest.org/).
For the former we did the following tests:
\begin{enumerate}
\item artificial data with zero background and no noise added,
\item artificial data with non-zero constant background, but no noise,
\item artificial data with non-zero constant background and corresponding
Poisson noise added.
\end{enumerate}
We found that our method could recover all objects that the
standard coadding gets, and then some. It fails if two or more objects
are blended such that they overlap by a large fraction.

For real data we obtained cutouts (image subsections) of size
60' x 2' for a few pointings on the sky at several different epochs.
Individual CCDs on Palomar-QUEST are
about 8'  wide. The dithering strategy followed
for the survey necessitates
choosing such narrow strips to ensure
that for the different scans we do not straddle different CCDs.
\begin{figure}[h] 
\begin{tabular}{lr}
{\includegraphics[height=1.005in]{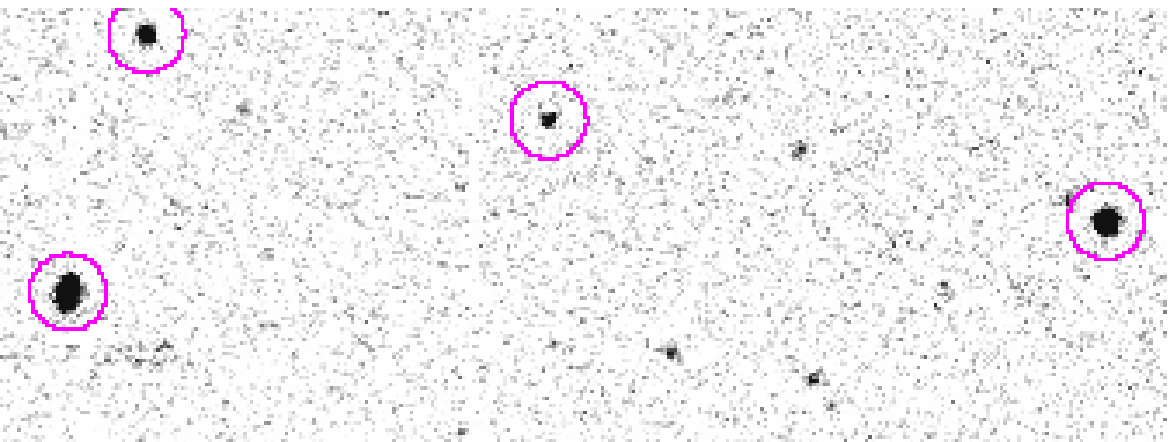} }&
{\includegraphics[height=.98in]{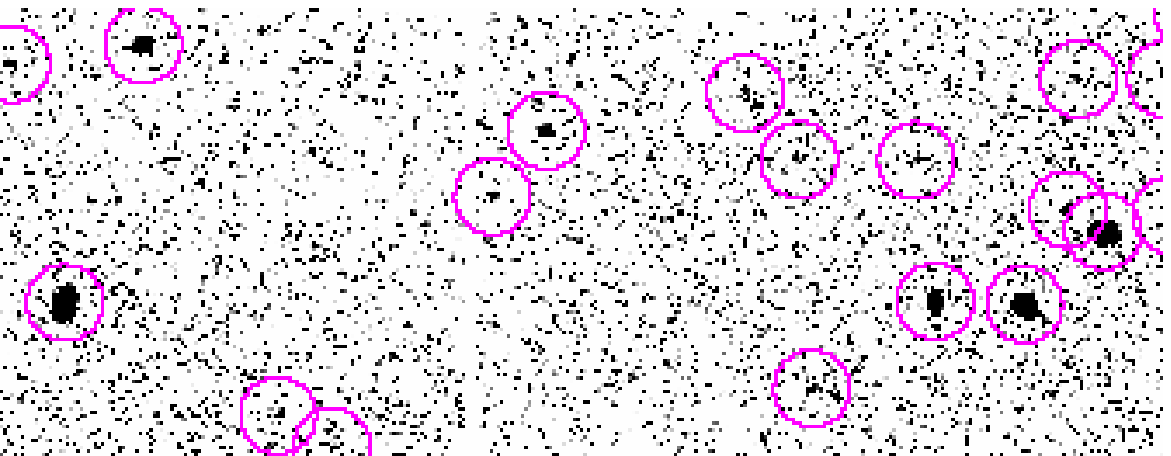}}\\
{\includegraphics[height=.965in]{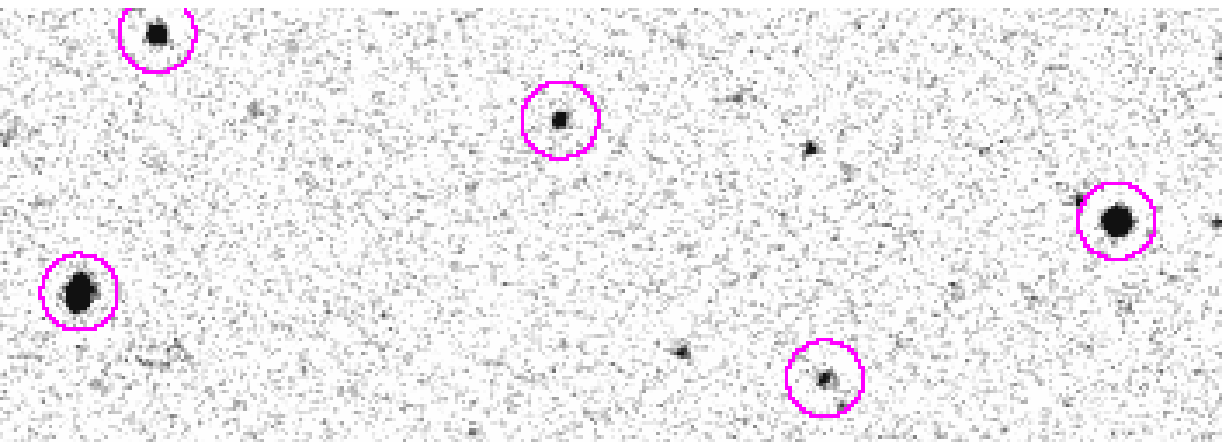}} &
{\includegraphics[height=.965in]{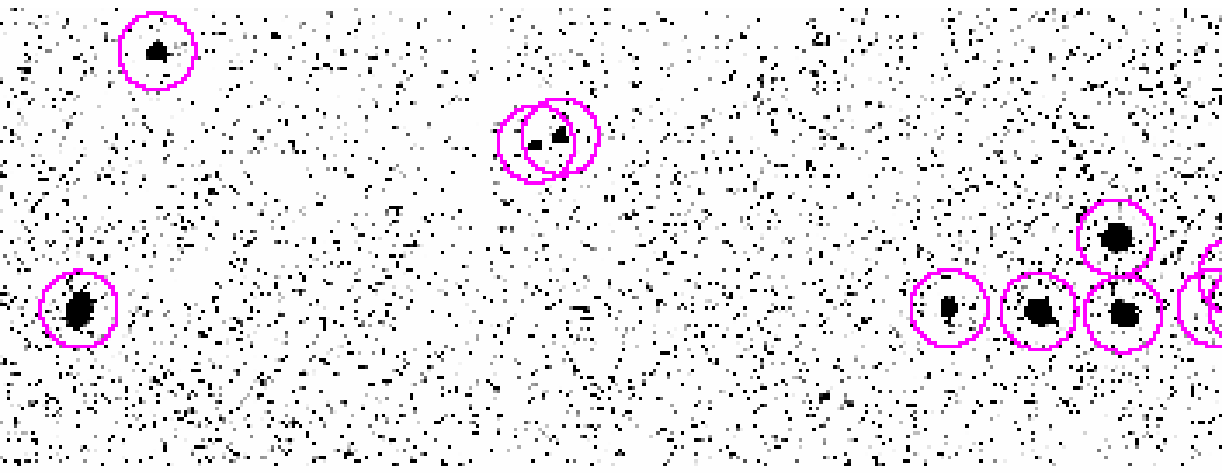}} \\
{\includegraphics[height=.99in]{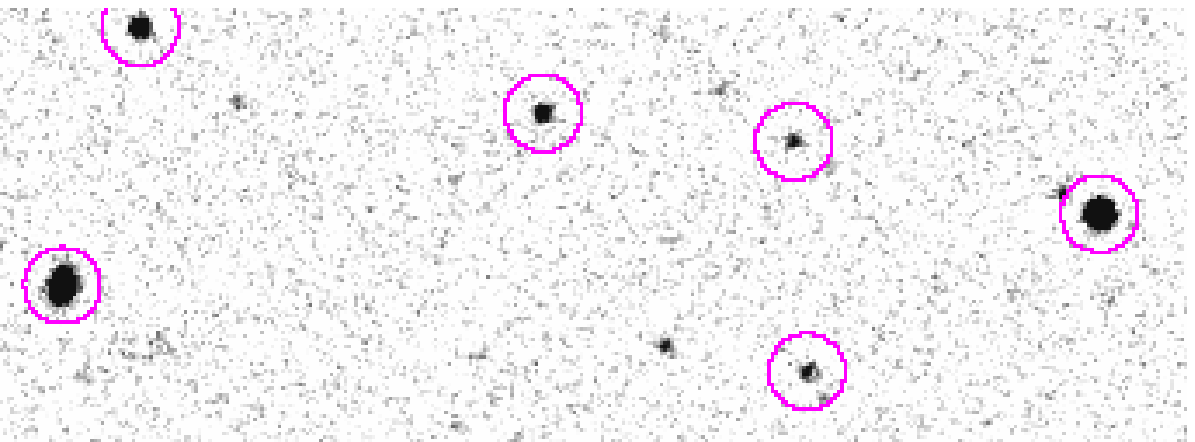} }&
{\includegraphics[height=.99in]{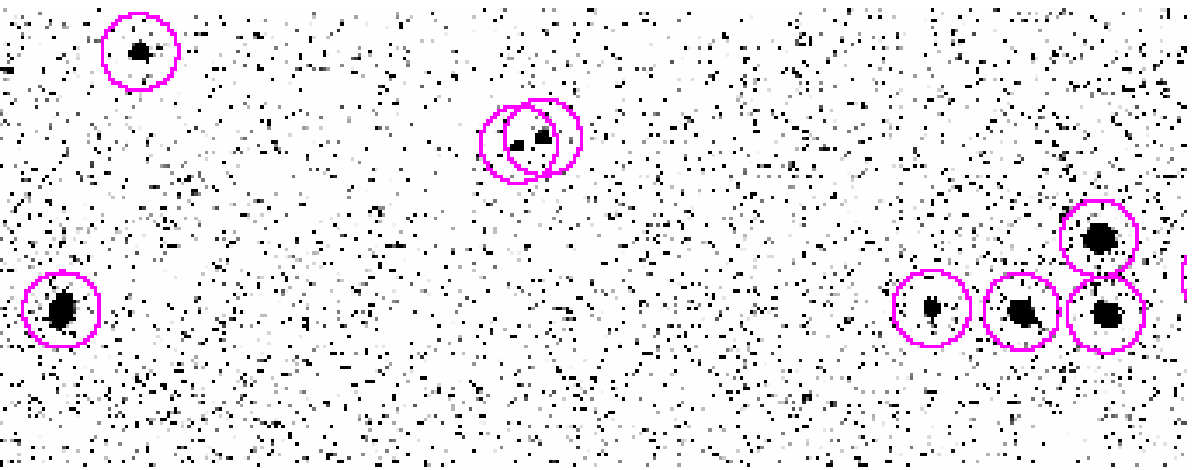}} \\
\end{tabular}
\caption{$B$ band images of a small portion of the sky from the
Palomar-QUEST multi-epoch survey.  The three panels show co-adds
from 3 (top), 5 (middle) and 7 (bottom) epochs.  The images on the left are
formed by a traditional pixel-by-pixel averaging procedure (mean
after rejection of minimum and maximum points) with4$\sigma$
sources circled. The images on the right are formed by the statistic
based on Mahalanobis
distance with 4$\sigma$ sources circled.}\label{fig.asteroids}
\end{figure}
(This does not mean that we can not do larger areas. It is just that for
current exploratory investigations we wish to keep inter-CCD calibrations
out of the picture).
The images were coadded in the standard way as well as using our method.
Object catalogs were then obtained for both sets at a series
of significance levels.
 
The results there too have been good. We coadded 3, 5 and 7 images
using both methods and found that our method does almost as well
as the traditional method in most cases and better in many others. This
has been quantified with a few tests and tabulated, but more stringent
tests are needed for large scale use. An example is to change the number of neighboring pixels considered to constitute the correlation matrix.
 
Some preliminary results based on Palomar-Quest images are shown in
Figure \ref{fig.asteroids}. Our images are more grainy in appearance
than the traditional co-added images due to the removal of
correlation between adjacent pixels.  When more image layers are
added, the graininess decreases. The emergence of 3 (two in the
top right panel) objects, in line at the bottom right corner of
the lower panels of the three images, absent in
the left panels is likely due to the passage of a minor planetary
body (asteroid, Kuiper Belt Object) through the field.

\section{Other possibilities}
\begin{figure}[h] 
\centerline{\includegraphics[height=3.5in]{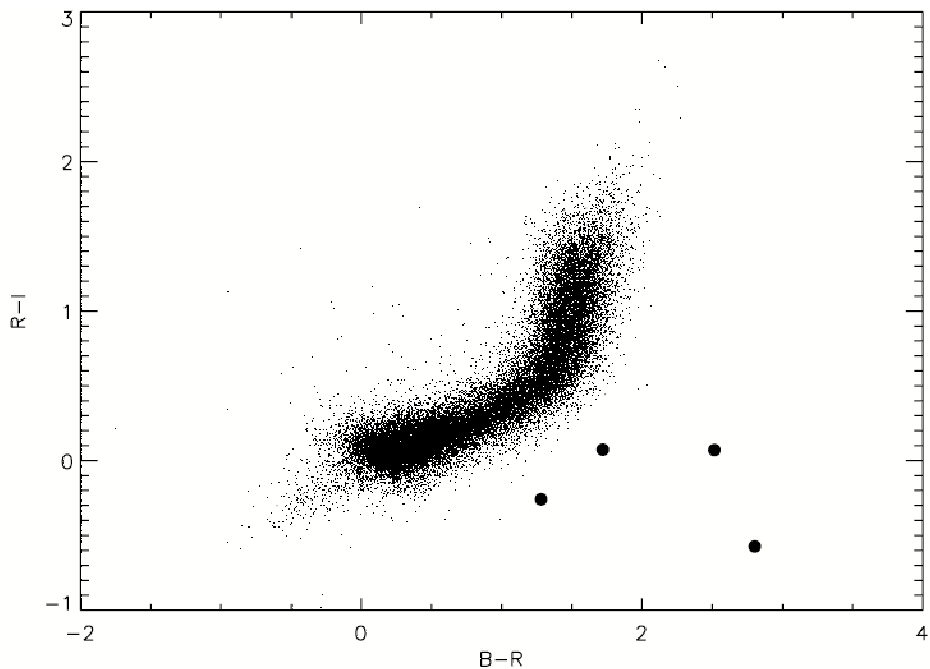}}
\caption{An example of high-redshift quasar selection, as outliers in a color-color parameter space, from a small area covered in the PQ survey.  Normal stars (dots) form a well-defined locus in this parameter space; quasar candidates (solid circles) deviate from this locus, while having exactly the same unresolved image morphology as the stars.}\label{fig.hi-zquasar}
\end{figure}
In addition to source
detection, a related problem that can be addressed with this technique
is that of identifying
faint objects in a particular region of color-color space.
A color indicates the ratio of fluxes in two bands. A color-color
diagram then is a plot which has such ratios formed from 3 bands:
A/B on x-axis, and B/C on y-axis. An example color-color diagram
depicting $B - R$ and $R - I$ colors from Palomar-QUEST is seen in Figure
\ref{fig.hi-zquasar}.
Objects of a given type occupy particular
sections of this plot. Thus for an area of interest
e.g. one that is expected to contain hi-z quasars,
one could try: $(f_k - f_i) * a*(f_i - f_z)$.
This can be tested by populating different colored objects and trying
detection in single band, combined bands and using the above technique.
Standard squaring as is used in calculating Euclidean distances may have
to be avoided as sign of the color is important. For bright
normal images this technique may not do well, but under lower
signal conditions it is expected to perform better.

\section{Future work}
 
As we refine the method for large data sets, we are starting to incorporate
it in to the Palomar-QUEST real-time pipeline. Initially some handholding will
be needed as Palomar-QUEST images in each of four filters for a given night
is about 1500 times the size of the test images tried so far.
The advantage will be to be
able to see fainter objects and hence curb the rate of
false alerts for transients.
The methodology will also set a  standard for the forthcoming synoptic
surveys like Pan-STARRS and LSST can reap rich benefits thereof.

\end{document}